\documentclass[
	letterpaper, 
	12pt, 
]{WhitePaperTemplate}

\reporttitle{ShieldUp!: A Case Study} 

\reportsubtitle{Inoculation Users Against Online Scams} 

\reportauthors{Abhishek Roy, Narsi G \& Sujata Mukherjee} 

\reportdate{February 2025} 

\rightheadercontent{\includegraphics[width=1.3cm]{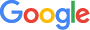}} 

\begin{document}


\thispagestyle{empty} 

\begin{fullwidth} 

	\vspace*{-0.075\textheight} 
	
	\includegraphics[width=3cm]{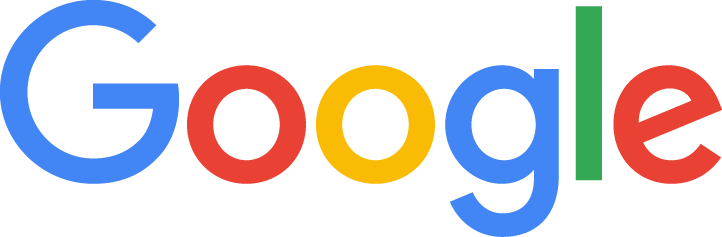} 

	\vspace{0.15\textheight} 
	
	{\fontsize{60pt}{62pt}\textbf{ShieldUp!}\par} 

	\vspace{0.05\textheight} 

	\parbox{0.9\fulltextwidth}{\fontsize{35pt}{37pt}\textbf{Inoculating Users Against Online Scams Using A Game Based Intervention}\par} 
	
    \vspace{0.03\textheight} 
	
	{\LARGE\textit{\textbf{A Case Study from India}}\par} 
	
	\vfill 
	
	{\Large\reportauthors\par} 
	
	\vfill\vfill\vfill 
	
	{\large\reportdate\par} 

\end{fullwidth}
\newpage


\thispagestyle{empty} 

\begin{twothirdswidth} 
	\footnotesize 
	
\section*{Acknowledgements}

The authors would like to express their sincere gratitude to Prof. Sander Van der Linden for his invaluable insights and guidance on the application of inoculation theory to scam prevention. We are also deeply indebted to Prof. Monica Whitty for her expertise on scam tactics and victim vulnerabilities, which significantly informed the development of ShieldUp!  We also extend our thanks to Shubham Goswami for his contributions to the research and his collaboration throughout this project.

\end{twothirdswidth}
\newpage


\begin{twothirdswidth} 
	\tableofcontents 
\end{twothirdswidth}

\newpage


\section{Introduction}

India is experiencing an unprecedented surge in online scams, fueled by rapid digitalization and widespread adoption of mobile payments. The scale and impact of these scams are staggering. According to the Indian Cyber Crime Coordination Centre (I4C), an average of 7,000 complaints were registered daily between January and April 2024, a 113\% increase over the same period in 2021–2023 ~\cite{Singh2024-dk}. Financial fraud accounts for three-quarters of all cybercrime in India, with approximately 47\% being UPI-based ~\cite{Tripathi2024-uf}. A 2023 YouGov survey found that 47\% of respondents said a friend or family member had lost money to an online scam ~\cite{Singh2023-qy}. \nonumsidenote{In the first four months of 2024, Indians lost over INR 1,750 Crore (approximately USD 209 million) to cybercriminals ~\cite{Tripathi2024-uf}.}

A recent global study revealed that Indian users receive an average of 12 scam messages daily, spending an estimated 1.8 hours per week dealing with them ~\cite{Dixit2023-vy}.  Moreover, in a 2023 McAfee survey, 60\% of Indian respondents reported difficulty identifying scam messages, attributing this to scammers' increasing use of AI ~\cite{Rao2023-ou}. India's diverse and rapidly evolving digital landscape, while empowering millions, has also become a fertile ground for scammers. The Unified Payments Interface (UPI), a cornerstone of India’s digital payment revolution, has unfortunately become a tool for fraudsters. This, coupled with the widespread use of mobile devices and varying levels of digital literacy, creates a complex challenge demanding innovative solutions.\sidenotequote[-1cm]{\textbf{\LARGE ``}They [scams] involve the misrepresentation of facts and the deliberate intent to deceive with the promise of goods, services, or other financial benefits that in fact do not exist or that were never intended to be provided.\\[4pt]~\cite{Titus1995-cf}}

Traditional awareness campaigns, like informational videos or text based tips, have unknown or unproven efficacy in empowering users to combat online scams ~\cite{Chugh2023-hg, Rahman2023-nl}. Furthermore, awareness campaigns frequently suffer from broad targeting, ill-defined success criteria, and lack of efficacy measurement (i:e change in user behavior), leading to uncertainty about their effectiveness. Research has also shown that simply providing more information or facts is unlikely to effectively change user behavior ~\cite{Christiano2017-jc}, particularly when scammers induce a "hot" emotional state, increasing likelihood of impulsive and error prone decision making ~\cite{Loewenstein2001-ch, Slovic2007-cm}.  Also, because scams rapidly evolve, training focused on specific scenarios risks obsolescence. Moreover, the rapidly evolving nature of online scams poses a challenge for traditional training methods that focus on specific scenarios. 

To address these limitations, we explored \textbf{game-based inoculation}, leveraging the psychological inoculation theory ~\cite{McGuire1961-yv}. This approach preemptively exposes individuals to weakened forms of persuasive messages (scam manipulation tactics) and provides preemptive refutation of these tactics so as to build resistance against future manipulation. Additionally, game-based learning offers a safe and engaging environment to experience and learn from potential threats, promoting active learning and knowledge retention ~\cite{Hu2023-ng,Breuer2010-ct}.
\nonumsidenote{As scammers constantly adapt their tactics and exploit new vulnerabilities, training that emphasizes recognizing static red flags may not adequately prepare users to identify novel scams, risking obsolescence and limited transferability of skills.}

This case study presents \textbf{ShieldUp!}, a mobile game prototype developed to inoculate users against common online scams prevalent in India. We detail the game's development—guided by research on scam tactics and user behavior—and highlight key findings from a pilot study. We discuss the implications of our findings for scam-resistant products and outline future research directions. 
\begin{marginfigure}
    \centering
    \includegraphics[width=\linewidth]{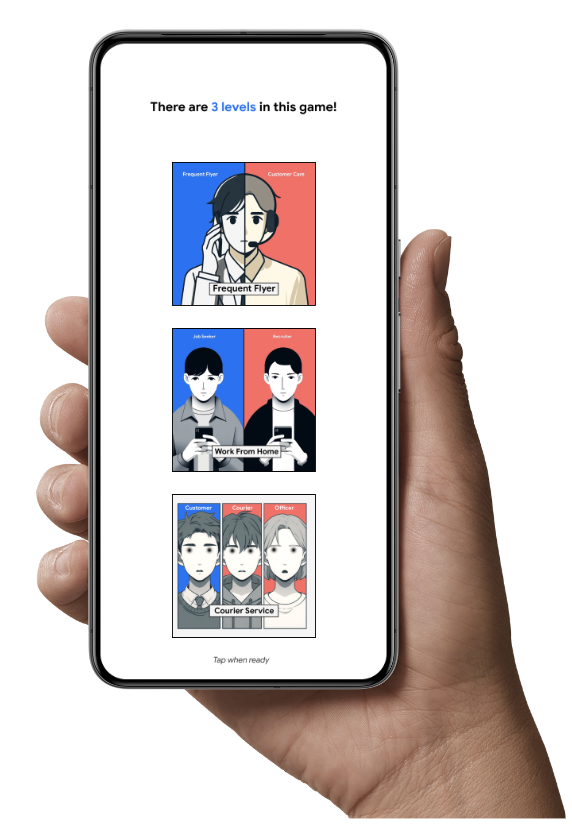}
    \caption{ShieldUp! has 3 increasing difficulty levels creating a skill ladder and cognitive scaffolding.}
    \label{fig:shieldUpLevels}
\end{marginfigure}

\section{Inoculation Theory \& Scam Prevention}

Traditionally, scam prevention strategies include general awareness campaigns, digital / financial literacy training, technology based solutions, and scam-specific training. Additionally, governments and regulatory institutions have intervened with legal and regulatory measures. However, a recent review of real-world fraud prevention interventions highlights a lack of rigorous evaluations and an over-reliance on diagnostic studies and untested assumptions ~\cite{Prenzler2019-ex}.

\textbf{General awareness campaigns}, such as UK's National Cyber Security Centre's Cyber Aware campaign and Australia's ACCC's Scamwatch, are designed to raise public awareness and serve as valuable repository of information for consumers~\cite{Ncsc-na, ScamWatch-lq}. These campaigns often use mass media, social media, and public service announcements. While they are prominently deployed as a tool by both the public ~\cite{Ncpi-bx} and private sectors~\cite{Chadha2022-gd}, there is limited direct evidence that they help raise awareness and are efficacious. Research consistently challenges the 'information deficit model,' showing that simply providing more information is unlikely to change behavior, particularly when other cognitive or emotional factors are at play \cite{Christiano2017-jc}. This is particularly relevant in scam contexts, where scammers often induce a 'hot' emotional state—fear, excitement, or greed—precisely to exploit these vulnerabilities in judgment and increase impulsive decisions \cite{Loewenstein2001-ch}. Combining these effects leads to reduced vigilance as victims fail to scrutinize information that might otherwise raise alarms. A meta-analysis of social engineering interventions found that awareness campaigns, while sometimes helpful, are generally less effective than more interactive and targeted approaches ~\cite{Bullee2020-hc}.
\begin{marginfigure}
    \centering
    \includegraphics[width=\linewidth]{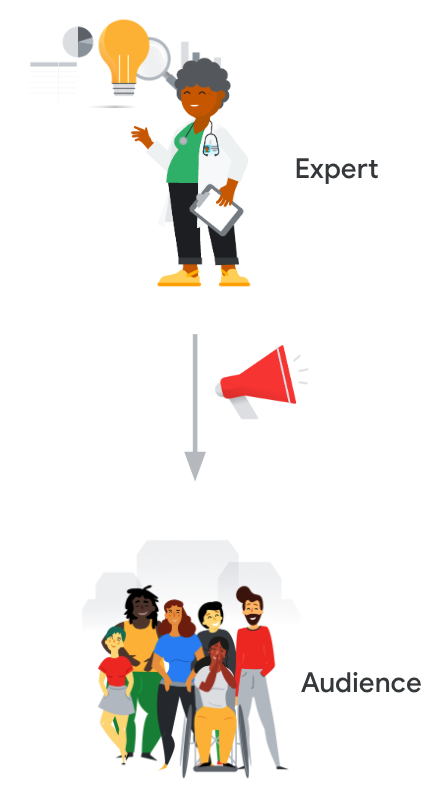}
    \caption{General Awareness Campaigns are based on Miller's 'information deficit model,' which assumes that users lack knowledge and that simply providing more information will change their behavior.~\cite{Miller1983-md}}
    \label{fig:genAwarenessModel}
\end{marginfigure}

\textbf{Digital literacy training} aims to equip individuals with critical thinking skills and knowledge to navigate the digital world safely. Organizations like MediaSmarts in Canada and Google's Be Internet Awesome program provide educational resources ~\cite{Jones2023-it, Hoechsmann2016-gl}.  This approach acknowledges that a lack of digital literacy can increase vulnerability. However, digital literacy training is often broad, and its effectiveness for scam prevention requires further research.

\textbf{Technology-based interventions} are crucial. Anti-phishing tools, spam filters, and multi-factor authentication prevent scams and protect accounts.  For example, the Singaporean government developed ScamShield that provides a product suite to help users check suspicious calls, websites, and messages ~\cite{ScamShield-ks}. While these offer a first line of defense, they cannot fully address the human element of scams, which often uses social engineering and psychological manipulation to bypass technological safeguards. Over-reliance on technology can create a false sense of security, discouraging critical thinking.

\textbf{Financial literacy education} can indirectly help prevent financial scams by empowering individuals to make informed decisions. ~\cite{Lusardi2014-fk} This approach emphasizes that financial knowledge can be a protective factor. For example, understanding the concept of compound interest and legitimate investment returns can help individuals spot unrealistic promises of quick riches, a common tactic in many scams. Burke et al. (2022) explored educational interventions, finding some positive effects ~\cite{Burke2022-wv}. However, maintaining and scaling this approach is challenging, and its impact on broader scam recognition is uncertain.

\subsection{Our Approach: Inoculation Against Manipulation Techniques}

Manipulation techniques, a perversion of persuasion techniques, are central to scams. While persuasion involves ethically influencing for mutual benefit; manipulation exploits victim's psychological vulnerabilities.  Robert Cialdini's \textit{The Psychology of Persuasion} outlines six core principles: reciprocity, scarcity, authority, consistency, liking, and social proof ~\cite{Cialdini1984-cs}. These principles leverage our natural tendencies for trust, social conformity, and reciprocation to increase compliance with requests, even when those requests are deceptive or harmful.  For example, scammers often create a false sense of scarcity or urgency to pressure victims into making quick decisions without careful consideration~\cite{Ariely2001-hq}.

\begin{marginfigure}
    \centering
    \includegraphics[width=\linewidth]{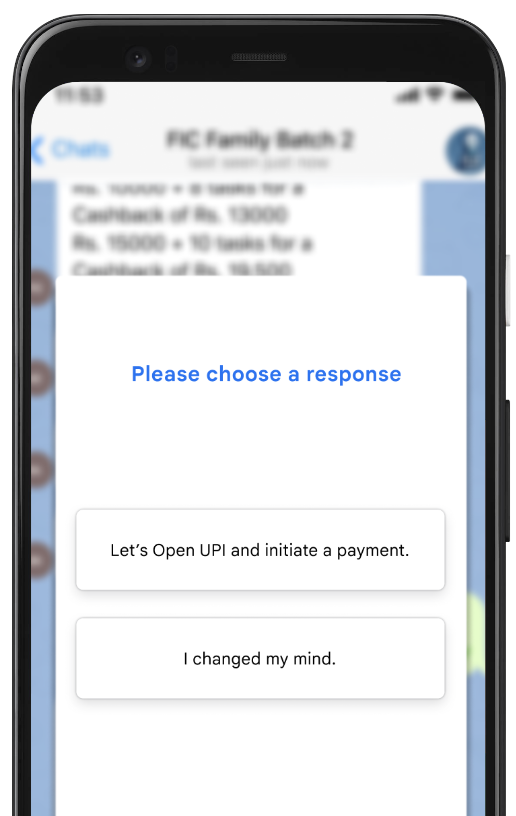}
    \caption{ShieldUp! confronts users to make choices simulating real-world scam scenarios}
    \label{fig:inGameDecision}
\end{marginfigure}

Since emotional appeals and psychological manipulation techniques often bypass rational decision-making~\cite{Petty1977-uv}, an approach that preemptively addresses these emotional vulnerabilities is crucial for effective scam prevention. Ergo, Psychological Inoculation, introducing users to weakened threats to increase resilience, is an approach worth exploring.

Introduced by McGuire (1961), inoculation theory posits that resilience can be built by introducing weakened attacks and then refuting them ~\cite{McGuire1961-yv}. An inoculation treatment includes a forewarning and preemptive refutation ~\cite{Compton2005-ms}. Thus, pre-exposing users to weakened scams and providing counterarguments could build future immunity. Recent studies have shown inoculation improves scam detection without harming trust ~\cite{Robb2023-gz, DeLiema2024-ds}.

Inoculation theory has demonstrated efficacy in bolstering attitudinal resilience against misinformation ~\cite{Roozenbeek2019-hi, McPhedran2023-zf}, extremist narratives~\cite{Saleh2023-fu}, and even in smoking prevention ~\cite{Pfau1992-st}. Inoculation theory, therefore, offers a robust framework for building resistance against persuasion and manipulation.

To summarize, therefore, our thesis was that inoculating users against common \textit{manipulation techniques} rather than specific scenarios might be a robust alternative to traditional methods of interventions against scams. We posit that this approach has the following advantages:

\begin{itemize}
    \item \textbf{Scalability:} Focusing on core techniques covers a wide array of scams.
    \item \textbf{Adaptability:} New techniques can be added easily.
    \item \textbf{Cross-protection:} Recognizing techniques in one context aids identification in new scenarios.
    \item \textbf{Cognitive Efficiency:} Teaching principles is more effective than teaching about specific instances.
    \item \textbf{Long-term Relevance:} Techniques persist even as specific scams evolve.
    \item \textbf{Empowerment:} Critical thinking skills extend beyond scams.
\end{itemize}
\begin{marginfigure}
    \centering
    \includegraphics[width=1.3\linewidth]{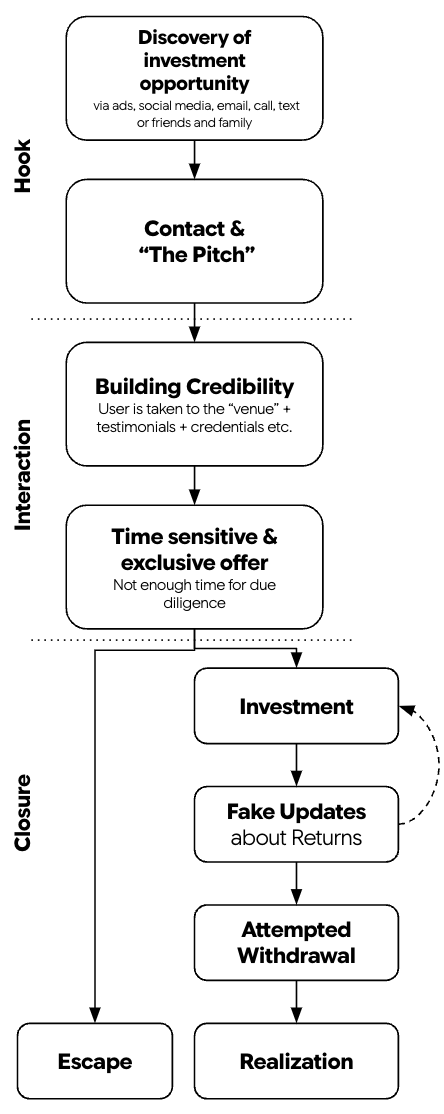} 
    \caption{Example of a user journey mapping exercise used to design a game scenario, illustrating the flow of interactions and decision points involved in a typical Investment Scam}
    \label{fig:investmentScamUJ}
\end{marginfigure}

\section{Developing ShieldUp!: A Game-Based Approach to Scam Prevention}

Building upon the principles of inoculation theory and a deep understanding of the Indian scam landscape, we developed ShieldUp!, a mobile game designed to empower users with the knowledge and skills to identify and avoid online scams.

\phantomsection
\label{sec:inoculation}
\subsection{Understanding Victim User Journeys}

Broadly, most scams follow a three phase trajectory: hook, interaction and closure. Hooks are designed to capture attention, followed by manipulation tactics to gain trust and exploit vulnerabilities, ultimately leading to loss of utility (monetary/information) for the victim. For example, a common tactic employed by scammers is to impersonate legitimate entities such as banks, courier services, or government agencies to gain the victim's trust ~\cite{Tripathi2024-uf}.  They may also leverage social pressure, using fake testimonials or creating a sense of urgency to coerce individuals into making hasty decisions.

Our preliminary analysis of India's scam landscape revealed that UPI scams, customer care scams, job scams, loan scams, marketplace scams, and crypto scams are among the most prevalent in India ~\cite{Tripathi2024-uf, Singh2024-dk}. 

We used Google advanced search features to search for a combination of keywords and identify user journey stories that were reported by 3rd party agencies like news and watchdog agencies or were self-reported by users on social media (Reddit, Twitter, Facebook etc.). We scanned and filtered through the first 5 pages of Google Search Results. In total, we collected 31 victim stories for 8 scam schemes, a full breakdown of which is available in \textit{Appendix A}.

We captured the following aspects of the Victim User Journeys (VUJs):
\begin{enumerate}
    \item Key aspects of a typical victim's user journey (Hook, Interaction \& Closure)
    \item Manipulation techniques and other tactics used by scammers in different parts of the VUJ.
    \item Victim vulnerabilities targeted
    \item Mental state and emotions felt by the victim (before, during and after the end of the VUJ).
\end{enumerate}

This analysis allowed us to identify common scam narratives, manipulation techniques employed by scammers, and vulnerabilities targeted in victims.

\subsection{Identifying Target Vulnerabilities}

Our analysis revealed a range of emotional and cognitive vulnerabilities that scammers frequently exploit:
\begin{marginfigure}
    \centering
    \includegraphics[width=1.3\linewidth]{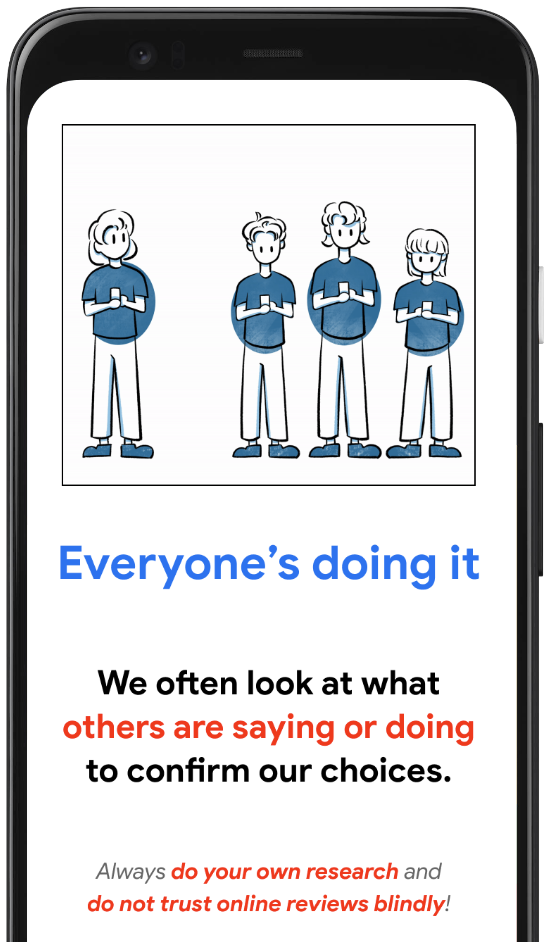}
    \caption{ShieldUp! has animated figures helping users understand manipulation techniques.}
    \label{fig:analyzeScam}
\end{marginfigure}

\begin{itemize}
    \item \textbf{Trust in Authority:}  Scammers often impersonate authority figures or well-known companies to leverage pre-existing trust. 
    \item \textbf{Desire for Quick Rewards:} Promises of easy money, high returns on investments, or quick solutions to problems often cloud judgment. 
    \item \textbf{Lack of Awareness:} Limited knowledge of common scam tactics, digital security measures, or the workings of specific online platforms can increase vulnerability.
    \item \textbf{Emotional Reasoning:} Scammers often induce fear, excitement, or a sense of urgency to trigger impulsive decision-making.
\end{itemize}

\subsection{Design Principles for Developing ShieldUp!}

Informed by the aforementioned research, we designed ShieldUp! as an interactive mobile game prototype that simulates real-world scam scenarios, allowing users to safely experience and learn from potential threats. 

Key design features of ShieldUp! include:

\begin{itemize}
    \item \textbf{Realistic Scenarios:} The game features a variety of scam scenarios based on real-world examples collected through our research, ensuring relevance and familiarity for Indian users.
    \item \textbf{Gradual Difficulty Progression (Skill Ladder):}  ShieldUp! utilizes a ``skill ladder'' approach, gradually increasing the complexity of scams as players progress through the levels. This allows users to build confidence and apply their growing knowledge to increasingly challenging situations. This approach incorporates principles of \textit{cognitive scaffolding}, where users are initially provided with more support and guidance, which is gradually reduced as their skills and understanding develop.
    \item \textbf{Interactive Storytelling:} The game utilizes an engaging narrative format, presenting players with choices that determine the course of the interaction. This interactive approach fosters active learning and allows users to experience the consequences of their decisions in a safe environment.
\begin{marginfigure}
    \centering
    \includegraphics[width=\linewidth]{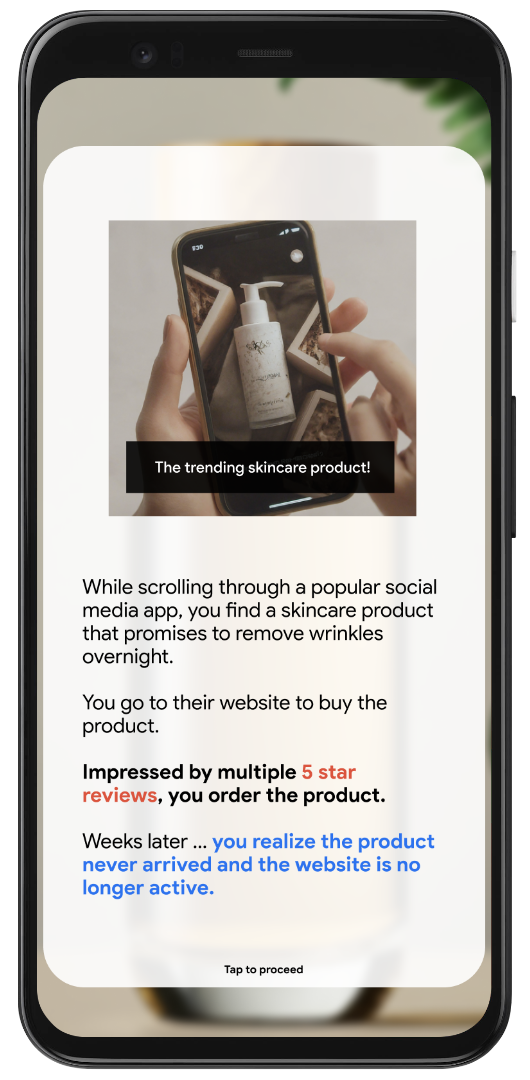}
    \caption{The game ends up with a short interactive quiz as an active recall exercise.}
    \label{fig:recallModule}
\end{marginfigure}
    \item \textbf{Explicit Teaching of Manipulation Tactics:}  After each scenario, the game provides clear explanations of the manipulation tactics used by the scammer, using simple language and memorable visuals.  This equips users with the knowledge and vocabulary to identify these tactics in real-world settings. 
    \item \textbf{Active Recall and Reinforcement:} ShieldUp! incorporates quizzes and interactive elements throughout the game play to reinforce learning and promote knowledge retention.
\end{itemize}

\paragraph{Targeting Specific Manipulation Tactics}

To ensure ShieldUp! effectively builds user resilience, we focused on incorporating the most prevalent and potent manipulation tactics identified in our research. These include:
\begin{itemize}
    \item \textbf{Social Proof (Conformity):} ShieldUp! presents scenarios where scammers leverage fake testimonials, inflated user numbers, or group chat dynamics to create a false sense of legitimacy and urgency.  The game then teaches players to recognize and question such attempts to manipulate their behavior through social pressure. ~\cite{Cialdini1999-jq}
    \item \textbf{Appeal to Authority:} The game features situations where scammers impersonate authority figures, such as bank officials or government representatives, to gain trust and coerce action.  ShieldUp! trains users to be wary of such authority claims and verify information through official channels. ~\cite{Milgram1965-cb}
    \item \textbf{Foot in the Door:}  ShieldUp! incorporates scenarios where scammers start with small, seemingly harmless requests that gradually escalate into larger demands. Players learn to recognize this pattern and set boundaries to prevent being drawn into exploitative situations. ~\cite{Freedman1966-sf}
\begin{marginfigure}
    \centering
    \includegraphics[width=\linewidth]{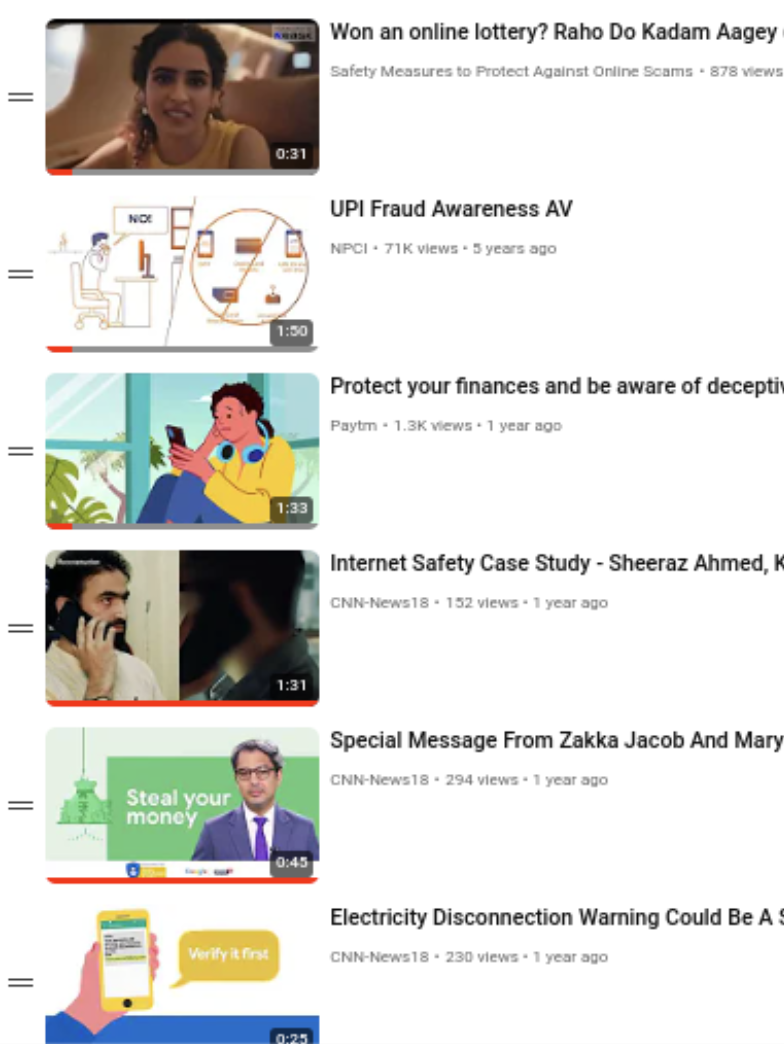}
    \caption{Active control group watched 5 short scam awareness campaign advertisements}
    \label{fig:genAwarenessPlaylist}
\end{marginfigure}
    \item \textbf{Urgency/Scarcity:} The game simulates scenarios where scammers use time pressure, limited-time offers, or impending deadlines to trigger impulsive decision-making.  ShieldUp! trains players to resist this pressure, step back, and assess situations rationally before taking any action.~\cite{Ariely2001-hq}
    \item \textbf{Appeal to Emotions:}  ShieldUp! includes scenarios designed to evoke strong emotions like fear, excitement, or greed, mirroring real-world tactics used to cloud judgment. The game teaches players to recognize these emotional triggers and seek alternative sources of information before making any decisions. ~\cite{Loewenstein2001-ch}
    \item \textbf{Norm Activation:} Activation of personal or social norms (what one believes they should do) can significantly motivate behavior, especially in contexts where these norms are made salient. The game simulates scenarios which try to trigger and help users recognize how norm activation is used in by scammers to manipulate them into their schemes. ~\cite{Schwartz1977-xy}
\end{itemize}

By directly addressing these specific manipulation tactics within the game's scenarios, ShieldUp! equips players with the knowledge and skills to recognize and resist such tactics in real-world settings.
Combining realistic scenarios, interactive gameplay, and explicit teaching of manipulation tactics, ShieldUp! aims to empower users with the critical thinking skills and behavioral strategies needed to navigate the digital landscape safely and confidently.

\section{Pilot Study and Key Findings}

To evaluate the efficacy of ShieldUp! in improving scam identification, we conducted a randomized controlled trial with 3,000 participants in India.

\subsection{Experimental Design}
The experiment had 4 parts: a pre-test to establish baseline scam and not scam identification abilities, the intervention, a post test to measure any change in abilities and a 21 day follow up assessment of ability. Participants were randomly assigned to one of three groups:
\begin{marginfigure}
    \centering
    \includegraphics[width=1.3\linewidth]{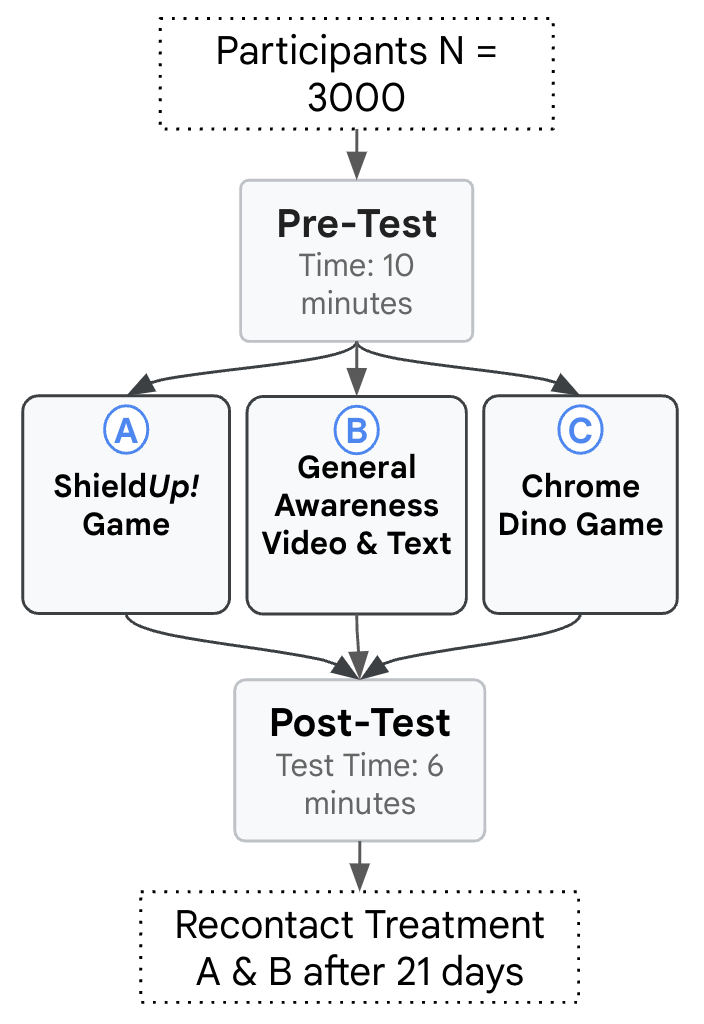}
    \caption{Experiment Design for the Randomized Controlled Trial}
    \label{fig:experimentSetup}
\end{marginfigure}

\begin{enumerate}
    \item \textbf{ShieldUp! Group (Treatment):} Participants in this group played ShieldUp! for approximately 15 minutes.
    \item \textbf{General Awareness Group (Active Control):}  Participants in this group watched a series of scam awareness videos and read safety tips for approximately 10 minutes.  The content for this group was curated from reputable sources, including government agencies, news channels and technology companies and can be found \href{https://www.youtube.com/watch?v=IoXWm8Cvp64&list=PLhu326--tQ14-hWmQ4WCXzhQ9pJuG-2J3}{\underline{here}} and the safety tips can be found in \textit{Appendix B}.
    \item \textbf{Chrome Dino Group (Control):}  Participants in this group played the Chrome Dino game for approximately 8 minutes. This group served as a baseline to control for the effects of simply engaging with a digital activity.
\end{enumerate}

\subsection{Measuring Scam Discernment Ability}

To assess a user's scam discernment ability, we developed the Scam Discernment Ability Test (SDAT-10), a 10-item test designed to measure an individual's capacity to identify and differentiate between scam and non-scam scenarios. The development of SDAT-10 involved a rigorous three-stage process: item generation, item condensation, and scale construction.

\paragraph{Item Generation} 

Drawing on I4C data, we identified the most prevalent scam types in India: Customer Care, Courier, Refund, Jobs, Friends/Family Impersonation, Online Shopping \& P2P Marketplace, and Crypto \& Investment scams ~\cite{Singh2024-dk}.  We then conducted a thematic analysis of 31 real-world scam victim user journeys reported in news media, social media, and civil society reports. This analysis allowed us to identify common scam narratives, manipulation techniques employed by scammers, and vulnerabilities targeted in victims.

Based on this analysis, we generated over 40 storylines, each with both scam and non-scam variants, resulting in a pool of over 80 potential test items.

\begin{marginfigure}
    \centering
    \includegraphics[width=\linewidth]{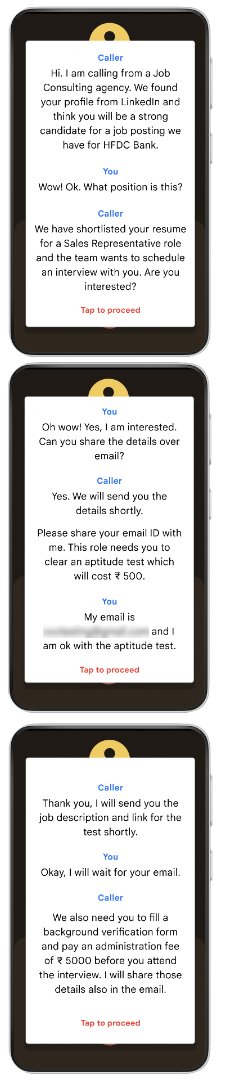}
    \caption{A SDAT-10 question item depicting a Job Scam}
    \label{fig:jobScam}
\end{marginfigure}

\paragraph{Item Condensation}

An expert committee reviewed the initial pool of items for content validity, resulting in the selection of 23 items for further testing. We tested these 23 items with a random sample of 360 users in India via an online survey. Based on Exploratory Factor Analysis (EFA), Item Response Theory (IRT) analysis, reliability analysis, and validity assessments, we shortlisted the 10 best performing items.

\paragraph{Scale Construction}

To mitigate memorization effects between pre- and post-tests, we created a second version (replica) of the SDAT-10 by making aesthetic changes to each item while maintaining the core storyline. Both versions were then tested again with a random sample of 600 users in India (300 per version) to confirm the 2 factor structure (one for scam identification ability and one for not scam identification ability) and to ensure that both the test sets were balanced.

The final SDAT-10 comprises five scam scenarios and five non-scam scenarios, each presented as a medium-fidelity interactive prototype depicting a typical scam user journey. Each scenario is accompanied by three questions assessing: 

\begin{enumerate}
    \item \textbf{Scam Compliance:} Likelihood of continuing engagement with the scenario.
    \item \textbf{Discernment:} Categorization of the interaction as "Scam" or "Not Scam."
    \item \textbf{Confidence:} Level of confidence in their assessment. 
\end{enumerate}

\subsection{Key Findings}

Using an ANCOVA model, we tested if there was an increase in user ability to correctly identify Scam and Not Scam scenarios post-intervention while controlling for their demographic profile (age, gender, income level, education level) and their pre-intervention ability of correctly identifying these scenarios. Since there were 5 scam and 5 not scam scenarios, their scam scores (1-5) denote their ability to identify scam scenarios correctly and their not scam scores denote their ability to identify not scam scenarios correctly.

\begin{figure}[h]
    \centering
    \includegraphics[width=1.3\textwidth]{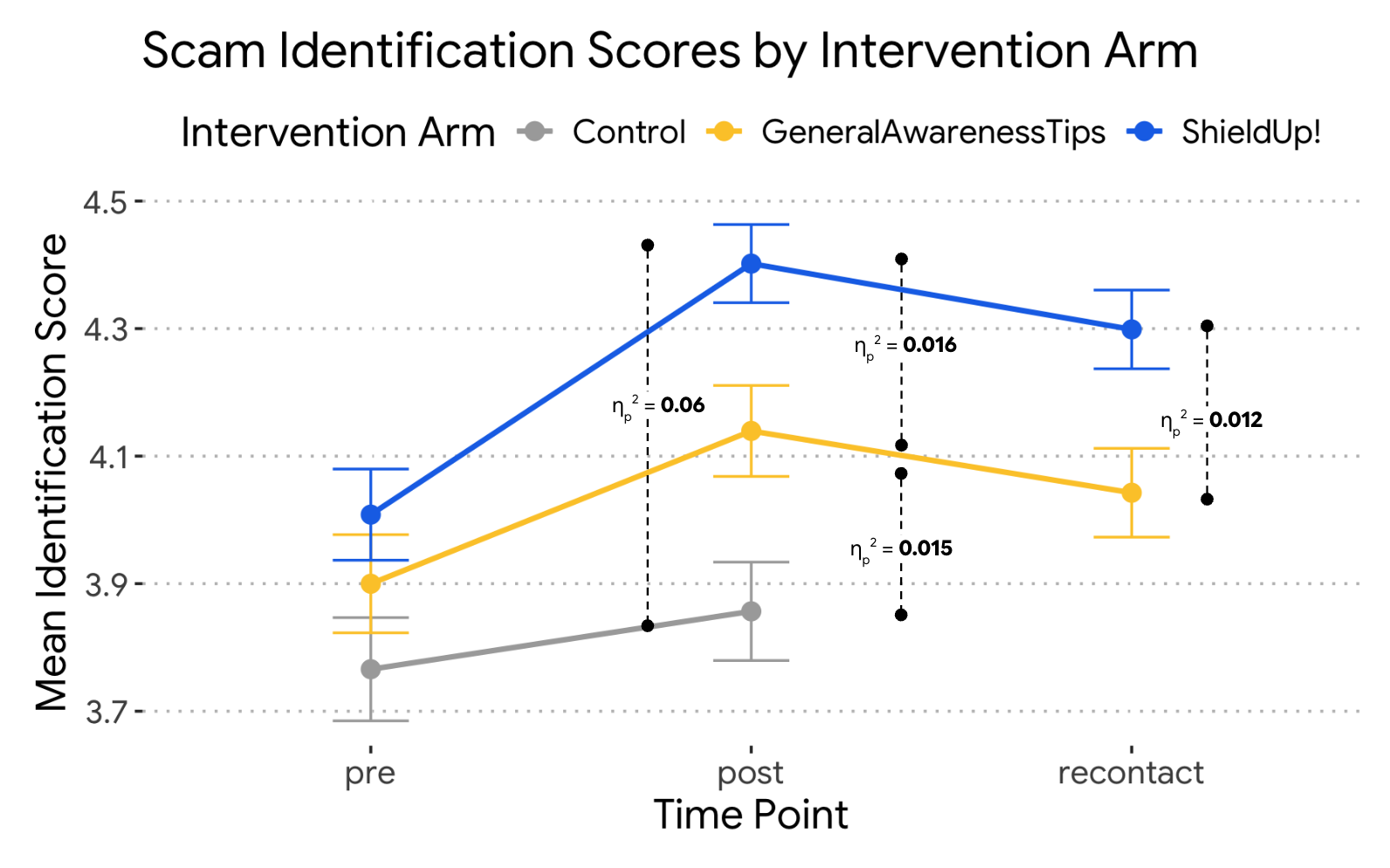}
    \caption{Mean Scam Identification scores by intervention arm at pre-test, post-test, and 21-day follow-up. Error bars represent standard errors.}
    \label{fig:scamIdentificationScores}
\end{figure}
\begin{figure}[h]
    \centering
    \includegraphics[width=1.3\textwidth]{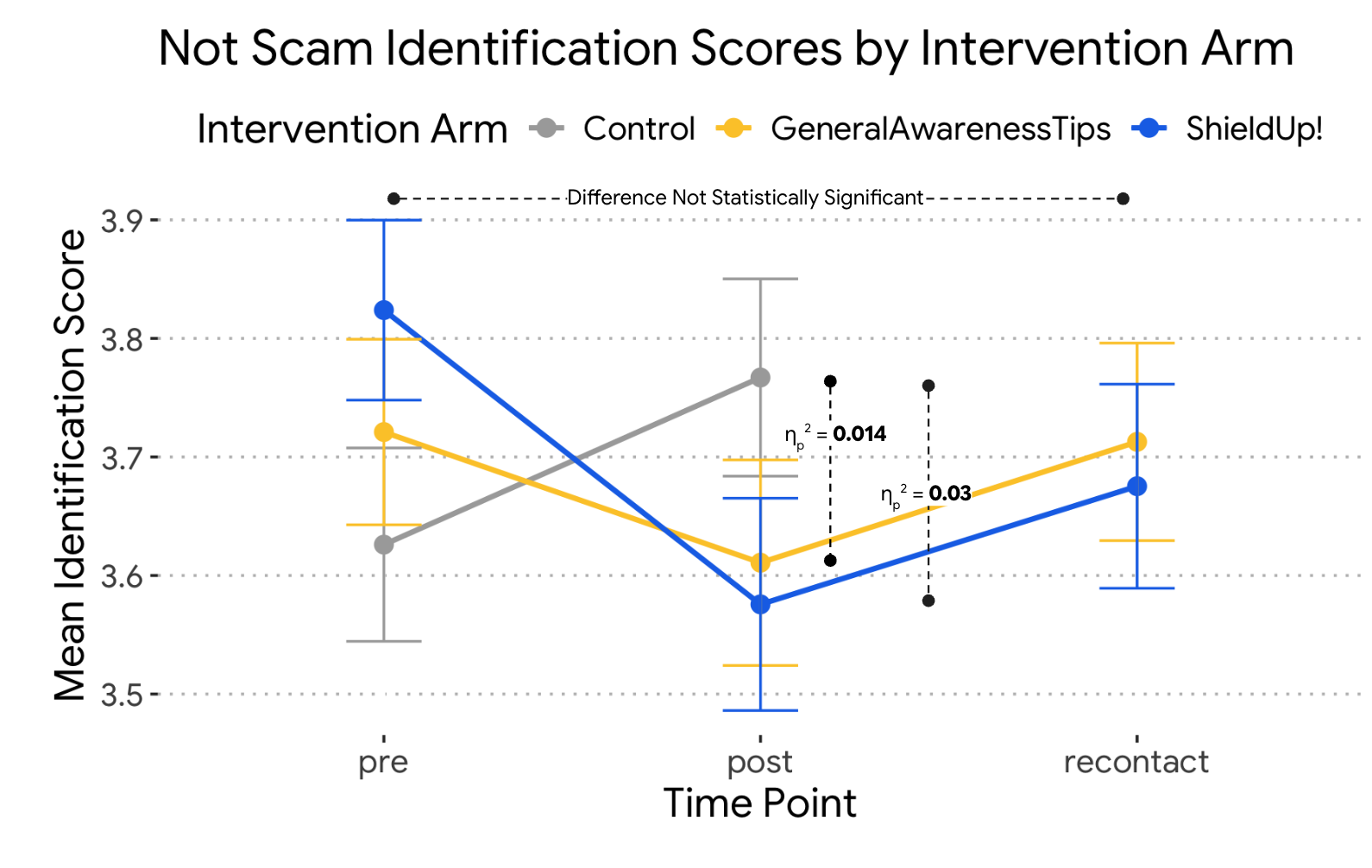}
    \caption{Mean Not Scam Identification scores by intervention arm at pre-test, post-test, and 21-day follow-up. Error bars represent standard errors.}
    \label{fig:notscamIdentificationScores}
\end{figure}

Our analysis revealed the following key findings:

\begin{enumerate}
    \item \textbf{ShieldUp! Improved Scam Identification Ability:}  Participants who played ShieldUp! demonstrated a statistically significant improvement in their ability to discern scam scenarios compared to both the general awareness and control groups.  This effect was observed immediately after the intervention and persisted at a follow-up assessment conducted 21 days later.
\nonumsidenote{Players experience an increased reluctance towards genuine offers after playing ShieldUp!, but this dissipates over time.}
    \item \textbf{Users have increased skepticism towards genuine offers following interventions, but this dissipates over time:} Immediately following both the ShieldUp! and general awareness interventions, participants exhibited heightened skepticism towards certain genuine online offers, such as cashback rewards and shopping discounts. This is similar to what Kubilay et al. observed in their intervention as well ~\cite{Kubilay2023-yi}. However, we observe that this initial increase in skepticism dissipated after 21 days for both ShieldUp! and General Awareness arms, suggesting that the interventions did not negatively impact trust in legitimate online interactions in the long term.
    \item \textbf{Effect Sizes:} The ShieldUp! intervention demonstrated a moderate effect size compared to the control group in improving scam identification. The general awareness intervention showed a smaller effect size compared to both the control and ShieldUp! groups.  This suggests that the active, game-based approach of ShieldUp! was more effective in enhancing scam identification than passive exposure to awareness materials.
\end{enumerate}
The findings of our pilot study provide promising evidence for the effectiveness of ShieldUp! as a game-based intervention for improving scam identification.  The game's ability to engage users, simulate realistic scenarios, and explicitly teach manipulation tactics appears to translate into tangible improvements in users' ability to identify and avoid potential scams. Moreover, the observed increase in skepticism towards genuine offers was temporary, suggesting that the game does not promote undue distrust in legitimate online interactions over longer periods of time while the improvements in scam identification abilities stay at an elevated level after 21 days. 

\section{Implications and Recommendations}
\begin{marginfigure}
    \centering
    \includegraphics[width=\linewidth]{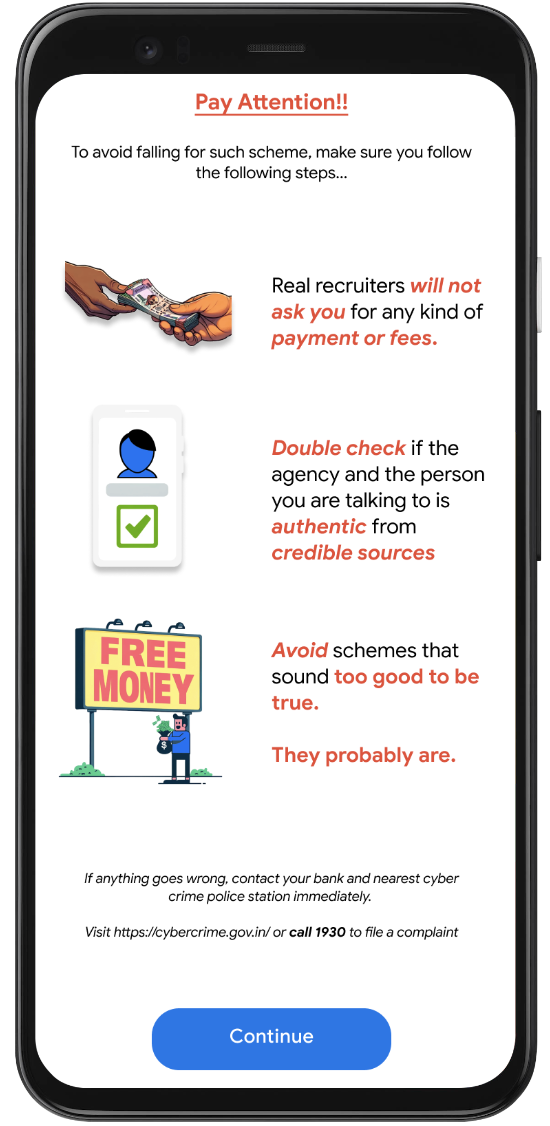}
    \caption{A safety advisory messages relevant to the scam scenario were shown after each level in the game}
    \label{fig:advisory_message}
\end{marginfigure}

The ShieldUp! pilot study demonstrated the effectiveness of inoculation-based gamified training in improving users' scam identification skills. This has significant implications for product design, policy, and future research in combating online scams.

\subsection{Implications for Product Design}

\begin{enumerate}
    \item \textbf{Integrate Inoculation-Based Training:}  
    \begin{itemize}
        \item \textbf{Interactive Learning Modules:} Introduce users to common manipulation techniques and provide opportunities to practice identifying scams in a safe environment. This can be achieved through engaging tutorials and interactive simulations that mimic real-world scam encounters, offering immediate feedback and guidance. These modules could be triggered at key moments in the user journey, such as when setting up a new account or making a transaction.
        \item \textbf{Just In Time Interventions:} Implement real-time warnings that alert users to potential red flags when engaging in risky online activities. These alerts should be context-specific and provide concise explanations of the manipulation tactics being employed. For instance, if a user receives a message containing a shortened link or a request for personal information, an alert could pop up explaining the potential risks.
    \end{itemize}

    \item \textbf{Context-Specific Interventions:} The effectiveness of inoculation interventions can be further enhanced by tailoring them to specific contexts, platforms, and demographics.
    \begin{itemize}
        \item \textbf{Platform Specificity:} Design interventions tailored to the specific platform where scams are prevalent. E-commerce sites could incorporate scam prevention modules during checkout, while social media platforms could provide tailored tips based on users' interaction patterns or offer warnings about suspicious profiles or messages.
        \item \textbf{Demographic Targeting:} Customize interventions based on user demographics, literacy levels, and risk profiles.
    \end{itemize}

    \item \textbf{Gamification for Enhanced Engagement:} ShieldUp!'s success highlights the value of gamified learning. Some best practices learnt from this experiment indicate that future efforts could benefit from incorporating design elements such as interactive narratives, effective sound design, reward mechanisms, cognitive scaffolding and active recall based quizzes to enhance user engagement and longer term knowledge retention. 
\end{enumerate}

\subsection{Policy Recommendations}

\begin{enumerate}
    \item \textbf{Promote Digital Literacy and Scam Prevention Education:} Policymakers can play a vital role in fostering a safer digital environment.
    \begin{itemize}
        \item \textbf{Curriculum Integration:} Integrate digital literacy and scam prevention education into school curricula at all levels. 
        \item \textbf{Public Awareness Campaigns:} Develop targeted public awareness campaigns using various channels to disseminate information about common scam tactics and prevention strategies.
        \item \textbf{Financial Literacy Programs:} Incorporate scam prevention education into financial literacy programs to help individuals understand the link between financial knowledge and scam susceptibility.
    \end{itemize}

    \item \textbf{Support Research and Evaluation:} Continued research is crucial. This includes investing in data collection and analysis to understand scam trends and evaluating intervention effectiveness. Further research should focus on developing and refining scam prevention strategies, particularly those addressing the psychological aspects of scam susceptibility.
\end{enumerate}

\subsection{Future Research Directions}

\begin{enumerate}
    \item \textbf{Longitudinal Studies:} Conducting longer time horizon (1-3 months) longitudinal studies can provide insights into the long-term effectiveness of interventions like ShieldUp! and determine whether booster interventions are needed to maintain heightened scam identification over time.

    \item \textbf{Cross-Cultural Adaptations:} Given the global nature of online scams, adapting ShieldUp! and similar interventions for diverse cultural and linguistic contexts is essential. Research should explore how to tailor game narratives, characters, and scenarios to resonate with different cultural values and beliefs.
\nonumsidenote{The exploration of personalized interventions holds great promise for optimizing scam prevention efforts. By tailoring interventions to individual needs and learning styles, we can potentially maximize their effectiveness and impact.}

    \item \textbf{Personalized Interventions:} To maximize effectiveness, future research should explore the development of personalized scam prevention interventions that take into account individual user characteristics, such as cognitive abilities, learning styles, risk preferences, and previous experience with scams. Personalized interventions can improve effectiveness by accounting for individual needs \cite{DeLiema2024-ds}. 
\end{enumerate}

\section{Conclusion}
Traditional awareness campaigns, while important, often fail to equip users with the skills and knowledge needed to effectively identify and avoid sophisticated scam tactics. This case study presents ShieldUp!, a mobile game prototype developed specifically for the Indian market, leveraging the principles of inoculation theory to provide users with a fun, engaging, and effective way to learn about online scams.  By simulating real-world scenarios, gradually increasing difficulty, and explicitly teaching manipulation tactics, we can empowers users to develop cognitive resistance and make safer choices online.

Our pilot study findings provide promising evidence that a game-based inoculation approach can significantly improve scam identification and potentially mitigate the growing threat of online scams. The observed improvements in scam identification, coupled with the dissipation of increased skepticism towards genuine offers over time, suggest that ShieldUp! can enhance user resilience without fostering undue distrust in legitimate online interactions. By embracing a proactive, evidence-based approach and fostering a culture of digital literacy and critical thinking, we can create a safer and more trustworthy digital world for everyone.



\newpage

\addcontentsline{toc}{section}{References} 

\begin{twothirdswidth} 
    \bibliography{bibliography} 

\begin{thebibliography}{}

\bibitem[Ncs, ]{Ncsc-na}
National cyber security centre uk.
\newblock \url{https://www.ncsc.gov.uk/}.
\newblock Accessed: 2024-10-15.

\bibitem[Ncp, ]{Ncpi-bx}
Ncpi fraud awareness program.
\newblock \url{https://www.npci.org.in/npci-in-news/knowledge-centre/fraud-awareness}.
\newblock Accessed: 2024-10-15.

\bibitem[Sca, ]{ScamShield-ks}
{ScamShield} website.
\newblock \url{https://scamshield.gov.sg/}.
\newblock Accessed: 2024-10-15.

\bibitem[Sca, 2024]{ScamWatch-lq}
 (2024).
\newblock Scamwatch australia.
\newblock \url{https://www.scamwatch.gov.au/}.
\newblock Accessed: 2024-10-15.

\bibitem[Ariely and Zakay, 2001]{Ariely2001-hq}
Ariely, D. and Zakay, D. (2001).
\newblock A timely account of the role of duration in decision making.
\newblock {\em Acta Psychol. (Amst.)}, 108(2):187--207.

\bibitem[Breuer and Bente, 2010]{Breuer2010-ct}
Breuer, J.~S. and Bente, G. (2010).
\newblock Why so serious? on the relation of serious games and learning.
\newblock {\em Eludamos: Journal for Computer Game Culture}, 4(1):7--24.

\bibitem[Bullee and Junger, 2020]{Bullee2020-hc}
Bullee, J.-W. and Junger, M. (2020).
\newblock How effective are social engineering interventions? a meta-analysis.
\newblock {\em Information and Computer Security}, ahead-of-print(ahead-of-print).

\bibitem[Burke et~al., 2022]{Burke2022-wv}
Burke, J., Kieffer, C., Mottola, G., and Perez-Arce, F. (2022).
\newblock Can educational interventions reduce susceptibility to financial fraud?
\newblock {\em J. Econ. Behav. Organ.}, 198:250--266.

\bibitem[Chadha, 2022]{Chadha2022-gd}
Chadha, S. (2022).
\newblock Inside google marketing: How we mobilized an industry-wide alliance in india to keep people safe online.
\newblock \url{https://www.thinkwithgoogle.com/intl/en-apac/future-of-marketing/management-and-culture/raising-cybersecurity-awareness-india/}.
\newblock Accessed: 2024-10-15.

\bibitem[Christiano and Neimand, 2017]{Christiano2017-jc}
Christiano, A. and Neimand, A. (2017).
\newblock Stop raising awareness already.

\bibitem[Chugh and Narang, 2023]{Chugh2023-hg}
Chugh, B. and Narang, L. (2023).
\newblock Are fraud awareness campaigns effective.
\newblock Technical report, Dvara Research.

\bibitem[Cialdini, 1984]{Cialdini1984-cs}
Cialdini, R. (1984).
\newblock Influence: Science and practice.

\bibitem[Cialdini et~al., 1999]{Cialdini1999-jq}
Cialdini, R.~B., Wosinska, W., Barrett, D.~W., Butner, J., and Gornik-Durose, M. (1999).
\newblock Compliance with a request in two cultures: The differential influence of social proof and commitment/consistency on collectivists and individualists.
\newblock {\em Pers. Soc. Psychol. Bull.}, 25(10):1242--1253.

\bibitem[Compton and Pfau, 2005]{Compton2005-ms}
Compton, J.~A. and Pfau, M.~W. (2005).
\newblock Inoculation theory of resistance to influence at maturity: Recent progress in theory development and application and suggestions for future research.
\newblock {\em Communication yearbook 29}.

\bibitem[DeLiema et~al., 2024]{DeLiema2024-ds}
DeLiema, M., Robb, C.~A., and Wendel, S. (2024).
\newblock What \textit{does} trust \textit{have} to do with it? training consumers to detect digital imposter scams.
\newblock {\em J. Financ. Crime}.

\bibitem[Dixit, 2023]{Dixit2023-vy}
Dixit, P. (2023).
\newblock An average indian gets 12 fake messages daily and these are the most common ones they fall prey to: Study.
\newblock \url{https://www.businesstoday.in/technology/news/story/an-average-indian-gets-12-fake-messages-daily-and-these-are-the-most-common-ones-they-fall-prey-to-study-405067-2023-11-08}.
\newblock Accessed: 2024-10-14.

\bibitem[Freedman and Fraser, 1966]{Freedman1966-sf}
Freedman, J.~L. and Fraser, S.~C. (1966).
\newblock Compliance without pressure: the foot-in-the-door technique.
\newblock {\em J. Pers. Soc. Psychol.}, 4(2):195--202.

\bibitem[Hoechsmann et~al., 2016]{Hoechsmann2016-gl}
Hoechsmann, M., DeWaard, H., and {Lakehead University / MediaSMarts} (2016).
\newblock {MAPPING} {DIGITAL} {LITERACY} {POLICY} {AND} {PRACTICE} {IN} {THE} {CANADIAN} {EDUCATION} {LANDSCAPE}.

\bibitem[Hu et~al., 2023]{Hu2023-ng}
Hu, B., Ju, X.-D., Liu, H.-H., Wu, H.-Q., Bi, C., and Lu, C. (2023).
\newblock Game-based inoculation versus graphic-based inoculation to combat misinformation: a randomized controlled trial.
\newblock {\em Cogn Res Princ Implic}, 8(1):49.

\bibitem[Jones et~al., 2023]{Jones2023-it}
Jones, L.~M., Mitchell, K.~J., and Beseler, C.~L. (2023).
\newblock The impact of youth digital citizenship education: Insights from a cluster randomized controlled trial outcome evaluation of the be internet awesome ({BIA}) curriculum.
\newblock {\em Contemp School Psychol}, pages 1--15.

\bibitem[Kubilay et~al., 2023]{Kubilay2023-yi}
Kubilay, E., Raiber, E., Spantig, L., Cahlíková, J., and Kaaria, L. (2023).
\newblock Can you spot a scam? measuring and improving scam identification ability.
\newblock {\em SSRN Electron. J.}

\bibitem[Loewenstein et~al., 2001]{Loewenstein2001-ch}
Loewenstein, G.~F., Weber, E.~U., Hsee, C.~K., and Welch, N. (2001).
\newblock Risk as feelings.
\newblock {\em Psychol. Bull.}, 127(2):267--286.

\bibitem[Lusardi and Mitchell, 2014]{Lusardi2014-fk}
Lusardi, A. and Mitchell, O.~S. (2014).
\newblock The economic importance of financial literacy: Theory and evidence.
\newblock {\em J. Econ. Lit.}, 52(1):5--44.

\bibitem[McGuire, 1961]{McGuire1961-yv}
McGuire, W.~J. (1961).
\newblock The effectiveness of supportive and refutational defenses in immunizing and restoring beliefs against persuasion.
\newblock {\em Sociometry}, 24(2):184--197.

\bibitem[McPhedran et~al., 2023]{McPhedran2023-zf}
McPhedran, R., Ratajczak, M., Mawby, M., King, E., Yang, Y., and Gold, N. (2023).
\newblock Psychological inoculation protects against the social media infodemic.
\newblock {\em Sci. Rep.}, 13(1):5780.

\bibitem[Milgram, 1965]{Milgram1965-cb}
Milgram, S. (1965).
\newblock Some conditions of obedience and disobedience to authority.
\newblock {\em Hum. Relat.}, 18(1):57--76.

\bibitem[Miller, 1983]{Miller1983-md}
Miller, J.~D. (1983).
\newblock Scientific literacy: A conceptual and empirical review.
\newblock {\em Daedalus}, 112(2):29--48.

\bibitem[Petty, 1977]{Petty1977-uv}
Petty, R. E. A.~D. (1977).
\newblock {\em A {COGNITIVE} {RESPONSE} {ANALYSIS} {OF} {THE} {TEMPORAL} {PERSISTENCE} {OF} {ATTITUDE} {CHANGES} {INDUCED} {BY} {PERSUASIVE} {COMMUNICATIONS}}.
\newblock PhD thesis.

\bibitem[Pfau et~al., 1992]{Pfau1992-st}
Pfau, M., Van~Bockern, S., and Kang, J.~G. (1992).
\newblock Use of inoculation to promote resistance to smoking initiation among adolescents.
\newblock {\em Commun. Monogr.}, 59(3):213--230.

\bibitem[Prenzler, 2019]{Prenzler2019-ex}
Prenzler, T. (2019).
\newblock What works in fraud prevention: a review of real-world intervention projects.
\newblock {\em Journal of Criminological Research, Policy and Practice}, 6(1):83--96.

\bibitem[Rahman et~al., 2023]{Rahman2023-nl}
Rahman, N. A.~A., Chong, C.~C., and Hisham, R. R. I.~R. (2023).
\newblock Money mule scams: Patterns of involvement and awareness campaign.
\newblock In {\em The European Proceedings of Social and Behavioural Sciences}, volume 132, pages 802--808. European Publisher.

\bibitem[Rao, 2023]{Rao2023-ou}
Rao, A.~K. (2023).
\newblock People in india are exposed to nearly 12 fake messages daily, reveals {McAfee’s} scam message study.
\newblock \url{https://www.thefinancialworld.com/people-in-india-are-exposed-to-nearly-12-fake-messages-daily-reveals-mcafees-scam-message-study/}.
\newblock Accessed: 2024-10-14.

\bibitem[Robb and Wendel, 2023]{Robb2023-gz}
Robb, C.~A. and Wendel, S. (2023).
\newblock Who can you trust? assessing vulnerability to digital imposter scams.
\newblock {\em J Consum Policy (Dordr)}, 46(1):27--51.

\bibitem[Roozenbeek and van~der Linden, 2019]{Roozenbeek2019-hi}
Roozenbeek, J. and van~der Linden, S. (2019).
\newblock Fake news game confers psychological resistance against online misinformation.
\newblock {\em Palgrave Communications}, 5(1):1--10.

\bibitem[Saleh et~al., 2023]{Saleh2023-fu}
Saleh, N., Makki, F., van~der Linden, S., and Roozenbeek, J. (2023).
\newblock Inoculating against extremist persuasion techniques – results from a randomised controlled trial in post-conflict areas in iraq.
\newblock {\em adv.in/psych}, 1(1).

\bibitem[Schwartz, 1977]{Schwartz1977-xy}
Schwartz, S.~H. (1977).
\newblock Normative influences on altruism.
\newblock In {\em Advances in Experimental Social Psychology}, volume~10 of {\em Advances in experimental social psychology}, pages 221--279. Elsevier.

\bibitem[Singh, 2023]{Singh2023-qy}
Singh, B. (2023).
\newblock A fifth of urban indians claim they have lost money due to a scam.
\newblock \url{https://business.yougov.com/content/48051-a-fifth-of-urban-indians-claim-they-have-lost-money-due-to-a-scam}.
\newblock Accessed: 2024-10-14.

\bibitem[Singh, 2024]{Singh2024-dk}
Singh, R. (2024).
\newblock Here is how much indians lost to cyber frauds between jan and apr of 2024.
\newblock \url{https://www.business-standard.com/india-news/here-is-how-much-indians-lost-to-cyber-frauds-between-jan-and-apr-of-2024-124052700151\_1.html}.
\newblock Accessed: 2024-10-14.

\bibitem[Slovic et~al., 2007]{Slovic2007-cm}
Slovic, P., Finucane, M.~L., Peters, E., and MacGregor, D.~G. (2007).
\newblock The affect heuristic.
\newblock {\em Eur. J. Oper. Res.}, 177(3):1333--1352.

\bibitem[Titus et~al., 1995]{Titus1995-cf}
Titus, R.~M., Heinzelmann, F., and Boyle, J.~M. (1995).
\newblock Victimization of persons by fraud.
\newblock {\em Crime Delinq.}, 41(1):54--72.

\bibitem[Tripathi, 2024]{Tripathi2024-uf}
Tripathi, R. (2024).
\newblock Indians lost over inr 1,750 crore to cyber fraud in first four months of 2024.
\newblock \url{https://economictimes.indiatimes.com/news/india/indians-lost-over-1750-crore-to-cyber-fraud-in-first-four-months-of-2024/articleshow/110444616.cms}.
\newblock Accessed: 2024-10-14.

\end{thebibliography}
    \bibliographystyle{apalike} 

\end{twothirdswidth}


\newpage

\section*{Appendices}

\begin{appendices}
\begin{fullwidth} 

\section{Appendix A: Thematic Analysis of Victim User Journeys}
\begin{longtable}{|p{3.5cm}|p{5.5cm}|c|p{5cm}|} 
\caption{Search Strategy and Results for Scam User Journey Artifacts}\\
\hline
\textbf{Scam Type} & \textbf{Search Keywords} & \textbf{Stories Analyzed} & \textbf{Example Story Link}\\
\hline
\endfirsthead
\caption{Search Strategy and Results for Scam User Journey Artifacts (Continued)} \\
\hline
\textbf{Scam Type} & \textbf{Search Keywords} & \textbf{Stories Analyzed} & \textbf{Example Story Link}\\
\hline
\endhead 
\hline 
\endfoot
Courier Scams & (“courier” OR “parcel” OR “fedex”) AND “scam” AND “india” & 5 &  \url{https://tinyurl.com/fedex-scam-story}\\
\hline
Customer Care Scams & (“customer care” OR “refund”) AND “scam” AND “india” & 5 & \url{https://tinyurl.com/customer-care-scam-story}\\
\hline
Jobs Scams & (“jobs” OR “recruitment”) AND “scam” AND “india” & 5 & \url{https://tinyurl.com/jobs-scam-story}\\
\hline
Friends/Family Impersonation Scams & (“friend” OR “family” OR “impersonat*”) AND “scam” AND “india” & 3 & \url{https://tinyurl.com/friends-scam-story}\\
\hline
Online Shopping \& P2P Marketplace Scams & ((“shopping” AND “online”) OR (“olx” OR “facebook marketplace” OR “quickr”)) AND “scam” AND “india” & 3 & \url{https://tinyurl.com/marketplace-scam-story}\\
\hline
Crypto/Investment Scams & “online” AND (“investment” OR “crypto” OR “cryptocurrency”) AND “scam” AND “india” & 4 & \url{https://tinyurl.com/inv-scam-story}\\
\hline
Loan Scams &  ("loan" OR "quick loan" OR "payday") AND ("scam" OR "fraud") AND "india" & 4 & \url{https://tinyurl.com/loan-scam-story}\\
\hline
Other UPI Scams &  ("qr code scam" OR "upi pin scam" OR "fake payment" OR "money transfer scam") AND "upi" AND "india" & 2 & \url{https://tinyurl.com/upi-scam-story}\\
\hline
\textbf{Total} & & \textbf{31} \\
\hline
\end{longtable}

\section{Appendix B: Safety Tips Infographic for the Active Control Group}
\includegraphics[]{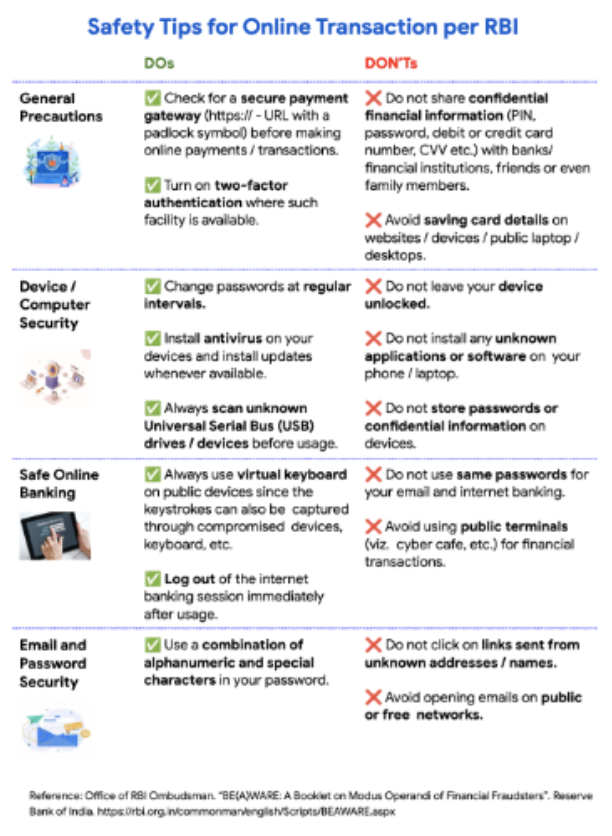}

\end{fullwidth} 
\end{appendices}

\end{document}